\documentstyle[12pt]{article}

\newcommand{\be}{\begin{equation}}
\newcommand{\ee}{\end{equation}}

\def\fun#1#2{\lower3.6pt\vbox{\baselineskip0pt\lineskip.9pt
\ialign{$\mathsurround=0pt#1\hfil##\hfil$\crcr#2\crcr\sim\crcr}}}
\title{\begin{flushright}
{\normalsize NUC-MINN-98/9-T\\
FAU-TP3-98/16\\ October 1998 \\
}
\end{flushright}
\vspace*{0.3in}
{\bf DISPERSION RELATION OF THE RHO-MESON
AT FINITE TEMPERATURE AND DENSITY}}
\author{{\bf V. L. Eletsky}$^{1,2}$
 \vspace*{0.2in} and {\bf J. I. Kapusta}$^3$\\
   {$^1$\it Institute of Theoretical and Experimental Physics}\\
   {\it B. Cheremushkinskaya 25}\\  \vspace*{0.2in}
   {\it Moscow 117218, Russia}\\
   {$^2$\it Institut f\"ur Theoretische Physik III}\\
   {\it Universit\"at Erlangen-N\"urnberg}\\  \vspace*{0.2in}
   {\it D-91058 Erlangen, Deutschland}\\
   {$^3$\it School of Physics and Astronomy}\\
   {\it University of Minnesota}\\
   {\it Minneapolis, MN 55455, USA}}

\date{}

\begin{document}

\maketitle
\begin{abstract}

The $\rho$ meson mass shift, width broadening, and spectral density
at finite temperature and nucleon density are
estimated using a general formula which relates the self-energy
to the real and imaginary parts of the forward scattering amplitude
on the constituents of the medium.  We saturate the scattering
amplitude at low energies with resonances, while at high energies
a Regge approach is taken in combination with vector meson dominance;
experimental data is used wherever possible.
The main result is that the $\rho$ meson becomes increasingly broad
with increasing nucleon density.  The spectral density is suppressed
in the resonance region $600 < M < 900$ MeV and enhanced
in the subresonance region $M < 600$ MeV.

\end{abstract}

\newpage

The problem of how the properties of hadrons change in hadronic or
nuclear matter in comparison to their free space values has attracted
a lot of attention.  Among the properties of immediate interest are the
in-medium particle's mass shift and width broadening.
Different models, as well as model independent
approaches, have been used to calculate these effects, both at finite
temperature and finite density.  It is
clear on physical grounds that the in-medium mass shift and width
broadening of a particle are only due to its interaction with the
constituents of the medium, for not too dense media anyway.
Thus one can use phenomenological information
on this interaction to calculate the mass shift and width
broadening: References \cite{sh,ei} give a few examples,
reference \cite{je} gives a relativistic field-theoretic derivation.

For meson $a$ scattering from hadron $b$ in the medium the
contribution to the self-energy is:
\begin{eqnarray}
\Pi_{ab}(E,p) &=& - 4\pi \int \frac{d^3k}{(2\pi)^3} \,
n_b(\omega) \, \frac{\sqrt{s}}{\omega}
 \, f_{ab}^{(\rm cm)}(s) \nonumber \\
&=& -\frac{1}{2\pi p} \int_{m_b}^{\infty} d\omega \,
n_b(\omega) \int_{s_-}^{s_+} ds \sqrt{s} f_{ab}^{(\rm cm)}(s)\, ,
\end{eqnarray}
where
$E$ and $p$ are the energy and momentum of the particle,
$\omega^2 = m_b^2 + k^2$,
\be
s_{\pm} = E^2 - p^2 +m_b^2 + 2(E\omega \pm pk) \, ,
\ee
$n_b$ is either a Bose-Einstein or Fermi-Dirac occupation number,
and $f_{ab}$ is the forward scattering amplitude.
The normalization of the amplitude corresponds to the standard form of the
optical theorem
\be
\sigma = \frac{4\pi}{q_{\rm cm}} {\rm Im} f^{(\rm cm)}(s) \, ,
\ee
where $q_{\rm cm}$ is the momentum in the cm frame.
The dispersion relation is determined by the poles of the propagator
after summing over all target species and including the vacuum
contribution to the self-energy:
\be
E^2 - m_a^2 - p^2 - \Pi_a^{\rm vac}(E,p) - \sum_b \Pi_{ab}(E,p) = 0 \, .
\ee
The applicability of eq. (1) is limited to those cases where interference
between sequential scatterings is negligible.

Taking various limits of eq. (1) is instructive. First of all, we
note that the cross section is invariant under longitudinal boosts.
It is convenient to know how the scattering amplitude transforms.
For the same relative velocity:
\begin{equation}
m_a f_{ab}^{(a\,{\rm rest\,frame})} = m_b f_{ab}^{(b\,{\rm rest\,frame})} =
\sqrt{s} f_{ab}^{(\rm cm)} \, .
\end{equation}
In the limit that the target particles $b$ move nonrelativistically
we can approximate $\omega$ in the first line of eq. (1) with
$m_b$, in which case
\begin{equation}
\Pi_{ab} = -4\pi f_{ab}^{(b\,{\rm rest\,frame})} \rho_b \, ,
\end{equation}
where $\rho_b$ is the spatial density.  Next consider the chiral
limit of pions serving as the target particles, relevant for
low temperature baryon free matter.  From eq. (5)
$\sqrt{s} f_{a \pi}^{(\rm cm)}=m_a f_{a \pi}^{(a\,{\rm rest\,frame})}$.
Since $f_{a \pi}^{(a\,{\rm rest\,frame})}$ involves two derivative
couplings of the pion to the massive state $a$ (Adler's theorem)
one sees from eq. (1) that $\Pi_{a\pi}\sim T^4$.  See also ref. \cite{T4}.
(In contrast note that $\Pi_{\pi\pi}\sim T^2$.)
Finally, if the self-energy is evaluated in the rest frame of $a$ it is
possible to do all the integrations but one.
\begin{equation}
\Pi_{ab}(E,p) = - \frac{m_a^2 T}{\pi p} \int_{m_b}^{\infty}
d\omega \ln\left[\frac{1-\exp(-\omega_+/T)}{1-\exp(-\omega_-/T)}
\right] f_{ab}^{(a\,{\rm rest\,frame})}(\omega)
\end{equation}
Here $\omega_{\pm} = (E \omega \pm pk)/m_a$.  This assumes that
$b$ is a boson; a similar formula ensues if it is a fermion.

In this paper we will estimate the $\rho$ meson dispersion relation
for finite temperature and baryon density for momenta up to a
GeV/c or so as this is very interesting for the production of dileptons
in high energy heavy ion collisions \cite{qm}.  Oftentimes such
investigations use low energy effective Lagrangians which are matched
to experimental data.  Here we will saturate the low energy part
of the scattering amplitude with resonances and use a combination of
vector meson dominance (VMD) and Regge theory at high energy.

We will assume that $\rho$-mesons are
formed during the last stage of the evolution of hadronic matter created in
a heavy ion collision, when the matter can be considered as
a weakly interacting gas of pions and nucleons.
This stage is formed when the local temperature is on the order of
100 to 150 MeV and when the local baryon density is on the order
of the normal nucleon density in a nucleus.
The description of nuclear matter as a
noninteracting gas of nucleons and pions, of course, cannot be considered as
a very good one, so it is clear from the beginning that our results may be
only semiquantitative. The main ingredients of our calculation are $\rho\pi$
and $\rho N$ forward scattering amplitudes and total cross sections.

As we mentioned already, the scattering amplitudes are saturated by
resonances at low energies.  At high energies we determine them
with the aid of VMD:  $\sigma_{\gamma N}$ is well known
experimentally, $\sigma_{\gamma \pi}$ can be found by the Regge approach.
Afterwards $Ref_{\rho N}$ and $Ref_{\rho \pi}$
are determined from dispersion relations. Since
VMD allows one to find only the cross sections of transversally polarized
$\rho$-mesons, we restrict ourselves to this case. As was shown in
\cite{ei}, for energies greater than 2 GeV the effects on longitudinal
$\rho$-mesons in nuclear matter are much smaller than for transverse ones.
At finite temperature they are comparable \cite{gk}.
Therefore, our results should be multiplied by a factor ranging from
2/3 to 1 for unpolarized $\rho$-mesons.  The actual construction of
the cross sections and scattering amplitudes was reported in
\cite{us} to which we refer the interested reader.

We will consider the momentum
$p$ to be real and evaluate the scattering amplitudes on-shell,
that is, evaluate the self-energy at $E = \sqrt{p^2+m_{\rho}^2}$.
In this case Eqs. (1) and (4) take the form
\be
E^2 = m_{\rho}^2 + p^2 +\Pi_{\rho}^{\rm vac} +
 \Pi_{\rho \pi}(p) + \Pi_{\rho N}(p) \, .
\ee
Since the self-energy has real and imaginary parts so does
$E(p) = E_R(p) -i \Gamma(p)/2$.  In the narrow width approximation
the dispersion relation is determined from
\begin{eqnarray}
E_R^2(p) &=& p^2+m_{\rho}^2+{\rm Re}\Pi_{\rho \pi}(p)
+ {\rm Re}\Pi_{\rho N}(p) \, , \nonumber \\
\Gamma(p) &=& - \left[ {\rm Im}\Pi_{\rho}^{\rm vac} +
{\rm Im}\Pi_{\rho \pi}(p)
+ {\rm Im}\Pi_{\rho N}(p) \right]/E_R(p) \, .
\end{eqnarray}
The width of the $\rho$-meson in vacuum, $\Gamma_{\rho}^{\rm vac}
= - {\rm Im}\Pi_{\rho}^{\rm vac}/m_{\rho}$, is 150 MeV.
We can also define a mass shift and optical potential in the
usual way.
\begin{eqnarray}
\Delta m_{\rho}(p) &=& \sqrt{m_{\rho}^2+{\rm Re}\Pi_{\rho \pi}(p)
+ {\rm Re}\Pi_{\rho N}(p)} - m_{\rho} \, , \nonumber \\
U(p) &=& E_R(p) - \sqrt{m_{\rho}^2 + p^2} \, .
\end{eqnarray}
We shall evaluate these for temperatures of 100 and 150 MeV and
nucleon densities of 0, 1 and 2 times normal nuclear matter density
(0.155 nucleons/fm$^3$).  This is accomplished by utilizing
a Fermi-Dirac distribution for nucleons.  The nucleon chemical
potentials are 745 and 820 MeV for densities of 1 and 2 times
normal at $T$ = 100 MeV, and 540 and 645 MeV for densities of 1
and 2 times normal at $T$ = 150 MeV.  Anti-nucleons are not included.

Here we would like to make a trivial point that is nevertheless
not discussed much in the literature.  In our first paper \cite{us}
we defined the width in the rest frame of the $\rho$-meson.
In this paper we define the width in the rest frame of the thermal
system.  The former definition is conventional and most useful in
particle physics; the latter definition is the usual one in statistical
and many-body physics, whether the system be nonrelativistic or
relativistic.  Either definition is equally valid.  For example,
consider the latter definition in the limit of a vanishingly
small density.  The width becomes $\Gamma(p) = (m_{\rho}/E)
\Gamma_{\rho}^{\rm vac}$ which decreases with increasing momentum.
This is just the time dilation effect and has nothing to do with
the $\rho$-meson moving through a many-particle system.

Figure 1 shows the mass shift at temperatures of 100 and 150 MeV
and for nucleon densities of 0, 1 and 2 times normal density.
The effect with pions alone is negligible (on the order of 1 MeV).
The main effect comes from nucleons.
The effective mass increases with nucleon density and with momentum,
but is almost independent of temperature.
At zero momentum the mass shift is about
15 MeV, reaching about 55 MeV at a
momentum of 1 GeV/c for 1 times nucleon density.
For 2 times nuclear density
these mass shifts are about 30 and 110 MeV, respectively.
These trends and numbers are roughly consistent with other analyses
\cite{Wambach,Weise,Friman}.

Similar trends occur in the $\rho$ meson potential as may
be seen in figure 2.  The potential is positive, is on the order
of tens of MeV, and increases with density.

Figure 3 shows the behavior of the $\rho$ meson width with
temperature, density, and momentum.  Once again pions have very little
effect.  The main effect comes from nucleons.  Contrary to $\rho$
mesons moving in vacuum or through a pure pion gas the width
remains roughly constant with momentum when nucleons are present.
The width is about 240 MeV at 1 times nuclear density and about
370 MeV at 2 times nuclear density.  This means that the $\rho$ meson
becomes a rather poorly defined excitation with increasing
nucleon density.

The rate of dilepton production is directly proportional to the
imaginary part of the photon self-energy \cite{mt,w} which is
itself proportional to the imaginary part of the $\rho$ meson
propagator because of vector meson dominance \cite{gs,gk}.
\begin{equation}
E_+ E_- \frac{dR}{d^3p_+ d^3p_-} \propto
\frac{-{\rm Im} \Pi_{\rho}}{[M^2 - m_{\rho}^2 - {\rm Re} \Pi_{\rho}]^2
+ [{\rm Im}\Pi_{\rho}]^2}
\end{equation}
The vacuum part of $\Pi$ can only depend on the invariant mass,
$M^2 = E^2 - p^2$, whereas the matter parts can depend on
$E$ and $p$ separately.  However, in the approximation we are
using the scattering amplutudes are of necessity evaluated on
the $\rho$ meson mass shell.  This means that the matter parts
only depend on $p$ because $M$ is fixed at $m_{\rho}$.  In particular,
the imaginary part of the matter contribution does not vary with
$M$.  Note that, in general, it need not vanish until the one pion
threshold is reached because it is a scattering process, not a decay.
The vacuum parts can be obtained from the Gounaris-Sakurai formula
\cite{gs,gk}.  This formula gives a very good description of the
pion electromagnetic form factor, as measured in $e^+e^-$ annihilation
\cite{ss}, up to 1 GeV apart from a small mixing with the $\omega$ meson
which we are ignoring in this paper.
\begin{eqnarray}
{\rm Re}\Pi_{\rho}^{\rm vac} &=& \frac{g_{\rho}^2 M^2}{48\pi^2}
\left[ \left(1-4m_{\pi}^2/M^2\right)^{3/2}
\ln\left|\frac{1+\sqrt{1-4m_{\pi}^2/M^2}}{1-\sqrt{1-4m_{\pi}^2/M^2}}
\right| +8m_{\pi}^2\left(\frac{1}{M^2}-\frac{1}{m_{\rho}^2}\right)
\right. \nonumber \\
&-&\left. 2\left(\frac{p_0}{\omega_0}\right)^3 \ln \left(\frac{\omega_0+p_0}
{m_{\pi}}\right) \right] \\
{\rm Im} \Pi_{\rho}^{\rm vac} &=& -\frac{g_{\rho}^2 M^2}{48\pi}
\left(1-4m_{\pi}^2/M^2\right)^{3/2}
\end{eqnarray}
Here $2\omega_0 = m_{\rho} = 2\sqrt{m_{\pi}^2+p_0^2}$.  The
vacuum width is $\Gamma_{\rho}^{\rm vac} = (g_{\rho}^2/48\pi)
m_{\rho} (p_0/\omega_0)^3$.

The imaginary part of the propagator is directly proportional
to the spectral density.  The former is plotted in figure 4 for
a pure pion gas at a temperature of 150 MeV and in figure 5
for pions and nucleons at a density of 1 times nuclear and a temperature
of 100 MeV.  These parameters are characteristic of the final stages
of a high energy heavy ion collision.  Pions alone have very little
effect on the spectral density even at such a high temperature.
The effect of nucleons, however, is dramatic.  The spectral density
is greatly broadened, so much so that the very idea of a $\rho$
meson may lose its meaning.

The above observations and remarks on the relative importance
of pions and nucleons may need to be re-examined when really
applying these calculations to heavy ion collisions.  The pions
may be overpopulated in phase space, compared to a thermal
Bose-Einstein distribution, and this could be modeled either
by introducing a chemical potential for pions or simply by
multiplication by an overall normalization factor.  Pions
would need to be enhanced by a substantial factor (5 or more)
to make a noticeable contribution at a density of 0.155 nucleons
per fm$^3$.

Recently preliminary data in Pb-Au collisions at 160 GeV$\cdot$A
have been presented \cite{data} where, in studying the
$e^+e^-$ mass spectrum, it was
found that the $\rho$-peak is absent at $k_T(e^+e^-)< 400$ MeV,
but reappears as a broad enhancement at $k_T(e^+e^-)> 400$ MeV.
This appears to be just the opposite of our findings.  However, our
calculations refer to the $\rho$ momentum relative to the local
rest frame of the matter.  In a heavy ion collision, the matter
generally flows outward from the central collision zone.  Therefore
a low momentum $\rho$ meson may actually be moving faster relative
to the outflowing matter than a higher momentum one.  No conclusion
can really be drawn without putting our results into a space-time
model of the evolution of matter in a heavy ion collision.

In summary, we have studied the properties of the neutral
$\rho$ meson in a finite temperature pion gas with and without
nucleons present.  Nucleons play the dominant role.  They provide
a generally positive potential for the $\rho$ mesons and
greatly increase their width.  The $\rho$ meson spectral density is so
broadened that the $\rho$ may lose its identity as a well defined
particle or resonance.  Our results are based on experimental
information on the scattering amplitudes and as such provide
a direct extrapolation from zero temperature and density to
nonzero values of both.  At sufficiently high energy density
the matter can no longer be described very well as a gas
of noninteracting pions and nucleons.  Nevertheless the trends
must be obeyed by any realistic calculations of the $\rho$
meson in-medium.  Applications to thermal and hydrodynamic models
of heavy ion collisions are under investigation.

\section*{Acknowledgments}

We are indebted to B. L. Ioffe for many valuable discussions.
This work was supported in part by CRDF grant RP2-132,
Schweizerischer National Fonds grant 7SUPJ048716, RFBR grant
97-02-16131, and the US Department of Energy grant DE-FG02-87ER40382.
V. L. E. acknowledges support of BMBF, Bonn, Germany.

\section*{Figure Captions}

\noindent
Figure 1:  The $\rho$ meson mass shift as a function of momentum.
The dashed lines represent $T$ = 100 MeV, the solid lines represent
150 MeV.  The density of nucleons is indicated.\\

\noindent
Figure 2:  The $\rho$ meson potential as a function of momentum.
The dashed lines represent $T$ = 100 MeV, the solid lines represent
150 MeV.  The density of nucleons is indicated.\\

\noindent
Figure 3:  The $\rho$ meson width as a function of momentum.
The dashed lines represent $T$ = 100 MeV, the solid lines represent
150 MeV.  The density of nucleons is indicated.\\

\noindent
Figure 4:  The imaginary part of the $\rho$ meson propagator as a
function of invariant mass for fixed values of momentum as indicated.
The temperature is 150 MeV and the matter is free of nucleons.\\

\noindent
Figure 5:  The imaginary part of the $\rho$ meson propagator as a
function of invariant mass for fixed values of momentum as indicated.
The temperature is 100 MeV and the nucleon density is the same
as in ordinary nuclei: 0.155 nucleons/fm$^3$.\\

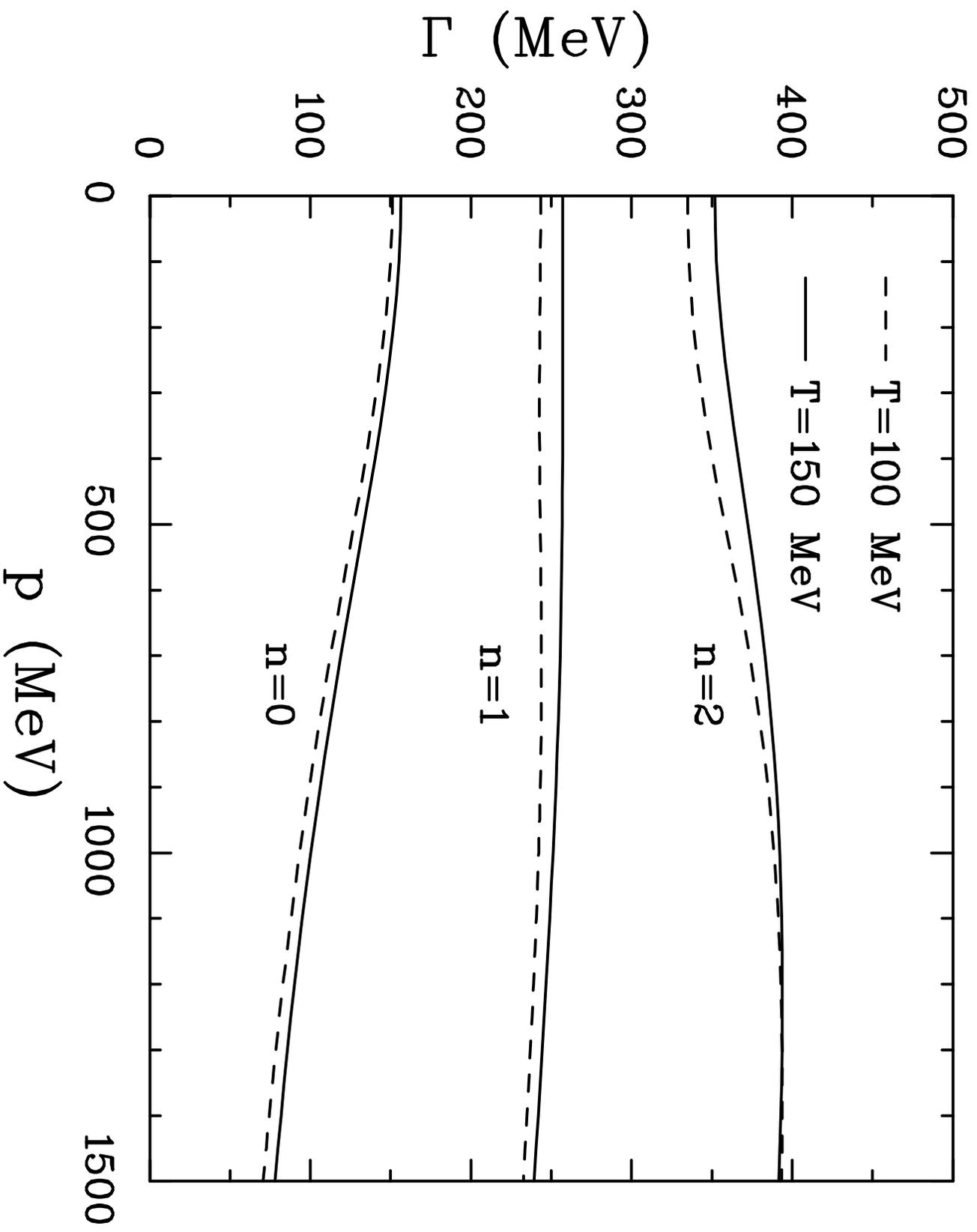
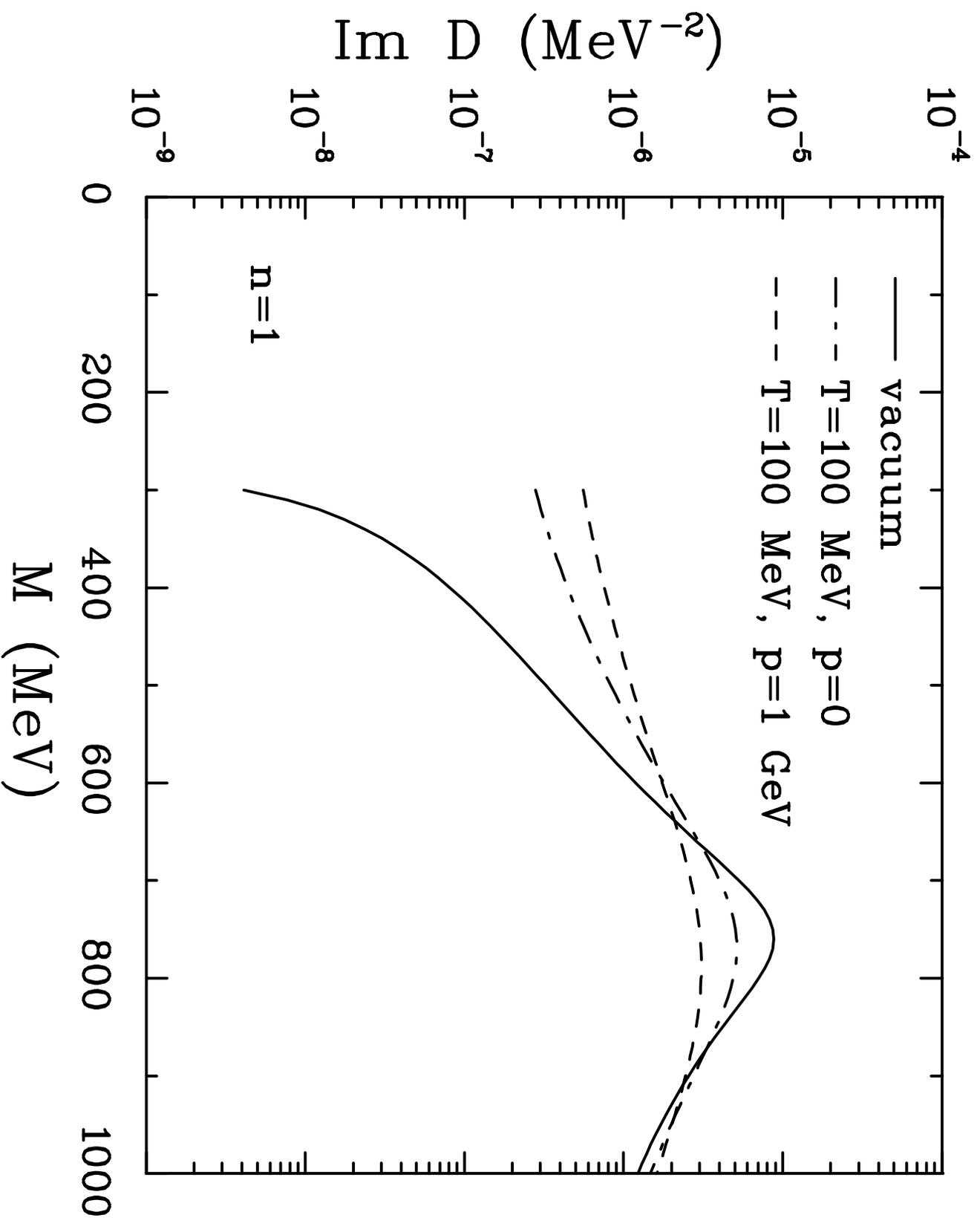
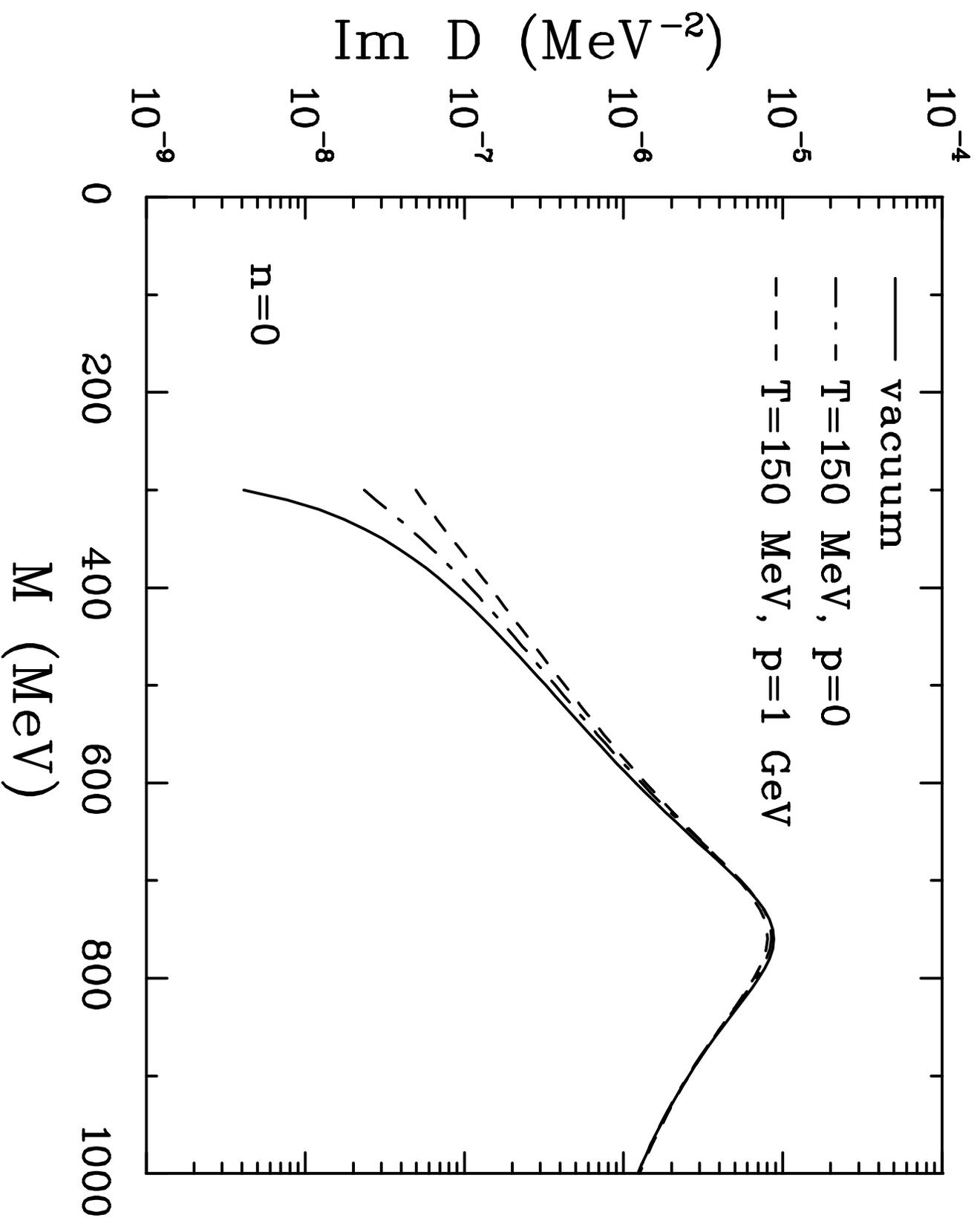
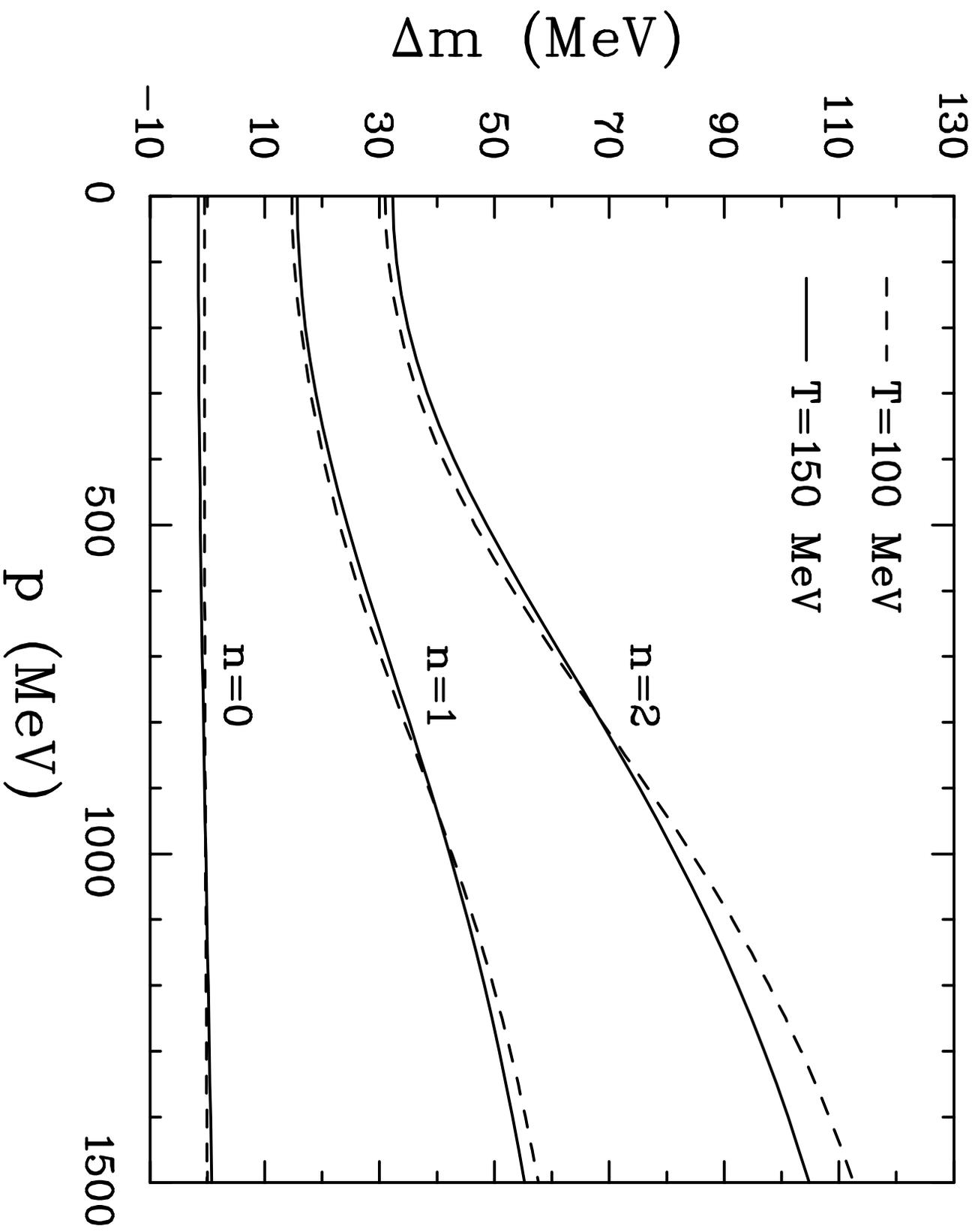
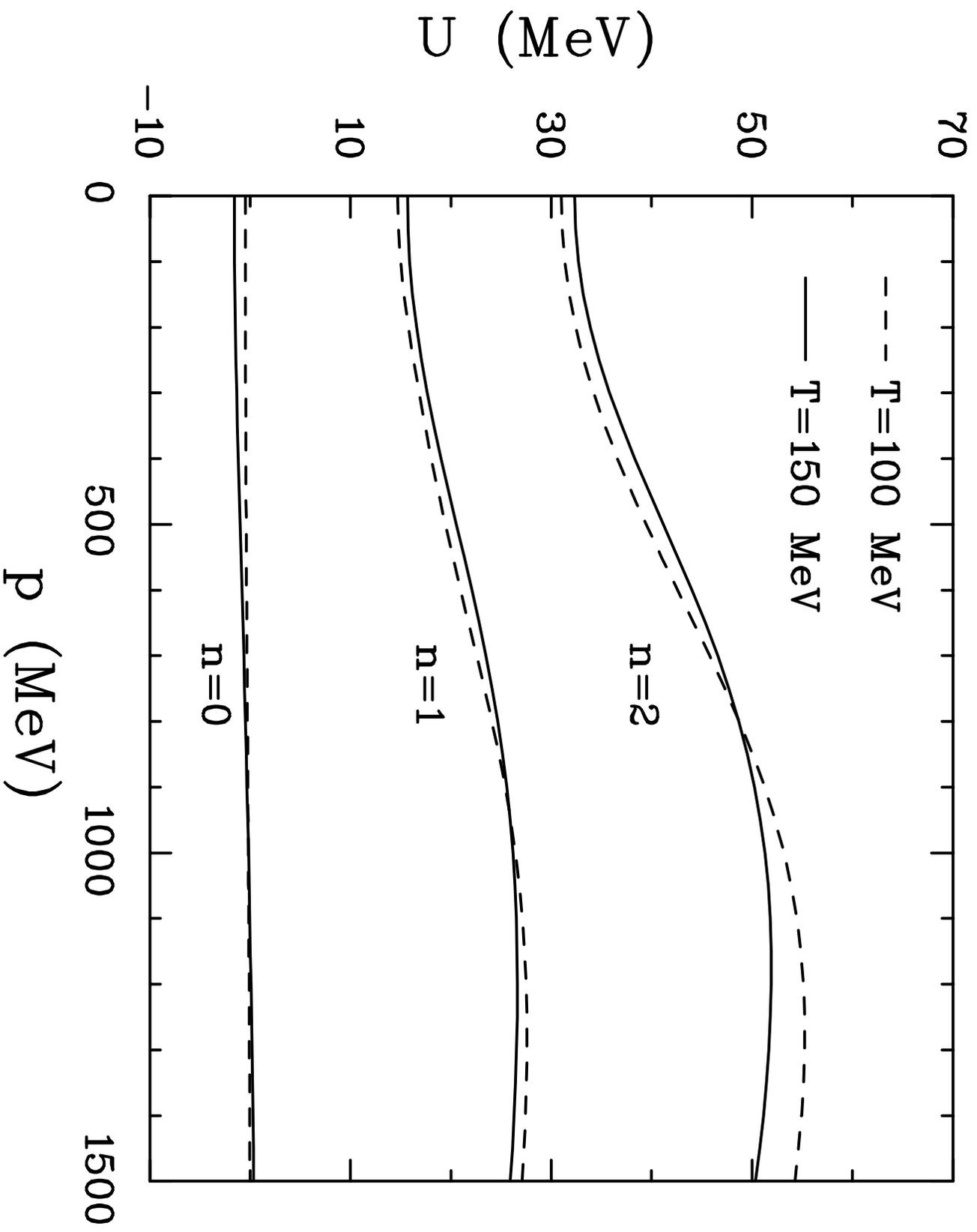
\end{document}